\documentclass{iau} 
\usepackage{graphicx}
\usepackage{xspace}
\usepackage[authoryear,round]{natbib}
\usepackage{biblio}

\def\newtopic{\par\vspace{5pt}\noindent}
\newcommand{\narreq}{\!=\!}
\newcommand{\narrsim}{\!\sim\!}

\newcommand{\nsimeq}{\!\simeq\!}
\newcommand{\nto}{\!-\!}

\newcommand{\nle}{\!\le\!}
\newcommand{\ngt}{\!\ge\!}
\newcommand{\nlt}{\!\le\!}
\def\gta{\;\lower 0.5ex\hbox{$\buildrel > \over \sim\ $}}
\def\lta{\;\lower 0.5ex\hbox{$\buildrel < \over \sim\ $}}

\def\kpc{\ifmmode {\> {\rm kpc}}\else {kpc}\fi}
\def\kms{\ifmmode {{\rm\ km\ s}^{-1}}\else {km s$^{-1}$}\fi}
\def\perkpc {\ifmmode {{\rm kpc}^{-1}}\else {\ kpc$^{-1}$}\fi}
\def\Msun{\ifmmode {\>M_\odot}\else {${\rm M}_\odot$}\fi}
\newcommand{\dg}{^\circ}

\newcommand{\pc}{\,\rm{pc}\xspace}
\newcommand{\kmskpc}{\,\rm{km\,s^{-1}\,kpc^{-1}}\xspace}

\newcommand{\Gyr}{\,\rm{Gyr}}

\title[Dynamics of the Galactic bulge and bar from surveys] %% give here short title %%
      {The barred inner Milky Way: dynamical models from surveys}

\author[Ortwin Gerhard]   %% give here short author list %%
{Ortwin Gerhard$^1$}
\affiliation{$^1$Max-Planck-Institute for Ex.~Physics, 
               Giessenbachstr.~1, D-85748 Garching, Germany \\
               email: {\tt gerhard@mpe.mpg.de}}
  
\pubyear{2017}
\volume{334}  %% insert here IAU Symposium No.
\jname{Rediscovering our Galaxy}
\editors{C. Chiappini, I. Minchev, E. Starkenburg \& M. Valentini, eds.}
\begin{document}

\maketitle

% TOTAL 0.4pp ie 10 line abstract. Total 44 lines/page.
\begin{abstract}
  The Milky Way is a barred galaxy whose central bulge has a box/peanut shape and consists
  of multiple stellar populations with different orbit distributions.  This review
  describes dynamical and chemo-dynamical equilibrium models for the Bulge, Bar, and inner
  Disk based on recent survey data. Some of the highlighted results include (i) stellar
  mass determinations for the different Galactic components, (ii) the need for a core in
  the dark matter distribution, (iii) a revised pattern speed putting corotation at
  $\sim6\kpc$, (iv) the strongly barred distribution of the metal-rich stars, and (v) the
  radially varying dynamics of the metal-poor stars which is that of a thick disk-bar
  outside $\sim\!1\kpc$, but changes into an inner centrally concentrated component with
  several possible origins. On-going and future surveys will refine this picture, making
  the Milky Way a unique case for studying how similar galaxies form and evolve.
\end{abstract}

\firstsection % if your document starts with a section,
\section{Overview: our barred Milky Way}

\par\noindent {\textsfbi{Introduction}}.  The stars in the Galactic bulge are mostly very
old ($\sim10\Gyr$), $\alpha$-enriched, and have a broad metallicity distribution (MDF),
pointing to an early rapid formation.  Yet starcounts have unambiguously established that
the bulk of the bulge stars are part of a box/peanut (B/P) bulge, and must therefore have
formed in the early Milky Way (MW) disk. The B/P bulge represents the inner 3D parts of the
Galactic bar, transiting into the planar long bar at about $2\nto3\kpc$ from the centre.
This confirms and corroborates long-standing evidence for a barred potential from NIR
photometry, early star counts, and non-circular gas motions in the bulge region
\citep[see][hereafter BHG16]{Rich2013,BHGerhard2016}.

Recently photometric and spectroscopic surveys have provided us with positions, velocities,
and metallicities for unprecedented samples of bulge stars.  These data have made it
possible to construct detailed dynamical models for the Bulge/Bar and its stellar
populations, providing new understanding of our MW, as will be discussed in this review.  In
the near future we expect great progress from the multiple surveys currently on-going or
planned, such as Gaia, VVV/X, DES, APOGEE, 4MOST or MOONS.

\newtopic {\textsfbi{Stellar masses and scale parameters}} The Galaxy is a luminous ($L_*$)
barred spiral with a stellar mass of $\sim5\times10^{10}\Msun$, an estimated circular
velocity at the Sun $V_0=238^{+5}_{-15}\kms$, and a relatively short and uncertain disk
scale-length, $R_d=2.4\pm0.5\kpc$ (BHG16).  From the dynamical models discussed below, the
corotation radius is $R_c=6.1\pm0.5\kpc$ (for $R_0=8.2\kpc$). The photometric stellar mass
of the Bulge and Bar is $M_{bb}=1.9\times10^{10}\Msun$, with the inner Disk adding
($R<5.3\kpc$) $M_{id}=1.3\times10^{10}\Msun$, both with uncertainty
$\sim0.1\times10^{10}\Msun$.  The stellar mass fraction of the Bulge, Bar, and inner Disk
together is thus $\sim65\%$, i.e., the major fraction of the MW's stars are in the inner
Galaxy.

\newtopic {\textsfbi{Galactic bulge density from starcounts}}.  The best structural
information for the Bulge comes from large samples of red clump giant (RCG) stars, for which
individual distances can be determined to $\sim\!10\%$. RCGs are representative for most of
the bulge stars, tracing old stellar populations within $10\%$ except for low metallicities
\citep{Salaris+Girardi02}. In the ARGOS survey they are prominent in the range of
metallicities [Fe/H]$\ge-1.0$ which contains $\sim\!95\%$ of their sample
\citep{Ness2013a}. RCGs have been used early-on as tracers of the bar asymmetry and have
been important in the discovery of the bulge X-shape \citep{McWilliam2010,Nataf2010}.  Using
$\sim\!8$ million RCGs from the VVV survey, \citet{Wegg+Gerhard13} measured the 3D bulge
density distribution in a box-shaped volume of $\pm 2.2 \times \pm 1.4 \times \pm 1.2\kpc$
(hereafter, the VVV box). They found a strongly barred ($\simeq1:2$) and peanut-shaped
density with bar angle $\phi_{\rm bar}\narreq27\dg\pm2\dg$. Along the bar axes, the central
($<\!1\kpc$) density distributions were found to be nearly exponential; in particular the
minor axis profile is exponential in the range $500\pc \nlt z \nlt 1.2\kpc$, with short
scale-length $z_0\narreq180\pc$, and shows no indication of a central $R^{1/4}$ component as
would be expected if the MW also hosted a substantial classical bulge.

Recently, large samples of Bulge RR Lyrae (RRL) stars have been identified in the OGLE and
VVV surveys.  RRL trace the very old, metal-poor population in the Bulge which is found not
to follow the barred RCG bulge \citep{Gran2016}.  From the number of RRL to RCG, $\sim1\%$
of the stars in the Bulge are in this old, more spheroidal distribution, which also rotates
less rapidly than the more metal-rich stars \citep{Kunder2016}.

\newtopic {\textsfbi{The Milky Way long bar from starcounts}}.  In N-body disk galaxy
models, B/P bulges are the inner 3D parts of a longer, planar bar and form through
buckling out of the galaxy plane and/or capture of stars by vertical resonances
\citep[e.g.,][]{Athanassoula2016}. B/P bulges in external galaxies are also embedded in
longer, thinner bars \citep{Erwin2013}, and observational evidence for buckling has
recenly been found by \citet{Erwin2016}.  The long bar in the Milky Way has been difficult
to characterize because of intervening dust extinction and the superposition with the
star-forming disk at low latitudes; see the summary in BHG16.  Using a density model for
RCGs from the combined 2MASS, UKIDDS and VVV surveys, \citet{Wegg2015} showed that the
MW's B/P bulge continuously transits outwards into a planar bar with half-length of
$5.0\pm0.2\kpc$ and bar angle $\phi_{\rm bar}=\sim28\dg-33\dg$ (for $R_0=8.3\kpc$),
consistent with the B/P bulge. Near the bar end the RCG overdensity is dominated by a
superthin component seen out to $l\nsimeq30\dg$; whereas the main, $180\pc$-scale-height
bar component reaches out to $R\nsimeq4.6\kpc$.  This region needs further study
\citep[e.g.,][]{Martinez-Valpuesta2011}.

\newtopic {\textsfbi{Bulge kinematics}}.  Data from the BRAVA survey for M-giant stars
showed that the B/P bulge rotates nearly cylindrically \citep{Kunder2012}, similar to B/P
bulges in external galaxies \citep{Molaeinezhad2016}. The near-cylindrical rotation is
seen for all metallicities up to $\rm{[Fe/H]}\narrsim-1$ in the ARGOS survey
\citep{Ness2013b}, and was comfirmed also at low latitudes $|b|\nlt2\dg$ by APOGEE
\citep{Ness2016}, and by GIBS \citep{Zoccali2017}. The overall velocity
dispersion ($\sigma_r$) profile decreases steeply with $|b|$. \citet{Shen2010} showed that
the cylindrical rotation of the Bulge could be fitted well by an N-body B/P bulge model,
but could not be fitted well if the MW contained a slowly rotating classical bulge with
more than 25\% of the bulge mass.

\section{Dynamical models for the Galactic bar}

\par\noindent {\textsfbi{Need for dynamical equilibrium models}}.  For interpreting star
counts in the Bulge and Bar a density model is sufficient; however, interpreting the
combined stellar positions and velocities requires a dynamical model. A dynamical model
describes the distribution of stars over orbits in the gravitational potential.  Even
though the Galactic bar will be evolving slowly (e.g., by angular momentum transfer), we
need to start with dynamical {\sl equilibrium} models in order to determine from the data
the current orbit distribution, and the current mass distribution and potential.  Later we
can perturb these models in order to study the evolution. In the bar region we have to
model a rapidly rotating triaxial system, but much of the essence of these complex models
is already contained in the simple spherical Jeans equation. This equation relates the
density, velocity dispersion, and orbit anisotropy of a tracer population with the total
dynamical mass and potential. In the rotating triaxial bar problem for the MW, we have a
more complicated geometry and model kinematic data for individual stars, but by
constructing a dynamical equilibrium model we again obtain relations between densities,
velocity moments, orbits, and potential, and find the best models by matching to the data.

\newtopic {\textsfbi{N-body model results}}.  N-body models have been very useful for
obtaining insight into the dynamics of B/P bulges and for illustrating possible origins of
the MW bulge. E.g., \citet{Abbott2017} showed that the B/P bulge is maintained by a wide
range of orbits, both resonant and non-resonant.  \citet{Gardner2014} explained the
different kinematics of stars on the near and far parts of the X in the Bulge.
\citet{Martinez-Valpuesta2013} showed that the Jacobi energies of stars are largely
conserved and, consequently, population gradients in the prior disk maintained during the
bar and buckling instabilities.  \citet{DiMatteo2014, Fragkoudi2017} and
\citet{Debattista2017} showed how stars on different orbits in the prior disk are mapped
into the final B/P bulge.  However, N-body models are not controllable and cannot
quantitatively match the multiple data points provided by large surveys (already many 1000s
but many more in future).

\newtopic {\textsfbi{Made-to-measure (M2M) particle models}}.  M2M models are well-suited
for incorporating large numbers of data constraints.  Their underlying principle is
simple. A suitable N-body model is constructed which captures the essence of the galaxy to
be modelled. The model is 'observed' just like the real galaxy is observed, including
survey selection functions (SSF). The model data are compared to the real data and the
difference is quantified in terms of a profit function. Then the weights (masses) of the
particles are modified such as to maximize the profit function with respect to the
weights. The modified N-Body model is then forward-integrated in time, and the cycle is
repeated until the model converges.  For the models described below, the NMAGIC
implementation of \citet{DeLorenzi2007, DeLorenzi2008} for observational constraints with
errors is used with various upgrades, such as for rotating potentials, on-the-fly
adaptation of the potential of the stars, and of the dark matter halo density, potential,
and particle distribution \citep[][herafter P17a]{Portail2017a}. The modelling starts with
controlled initial B/P bulge-bar-disk models adapted to specific shape and pattern speed,
and uses as data constraints the 3D bulge density, the RCG magnitude distributions in the
long bar, kinematics from the BRAVA, ARGOS, and OGLE surveys, as well as information on
the Galactic rotation curve.

\section{Bulge-Bar Dynamics: pattern speed, stellar, and dark matter mass distribution}

\begin{figure}[t]
%\vspace*{1.6 cm}
\begin{center}
  \includegraphics[width=0.46\linewidth]{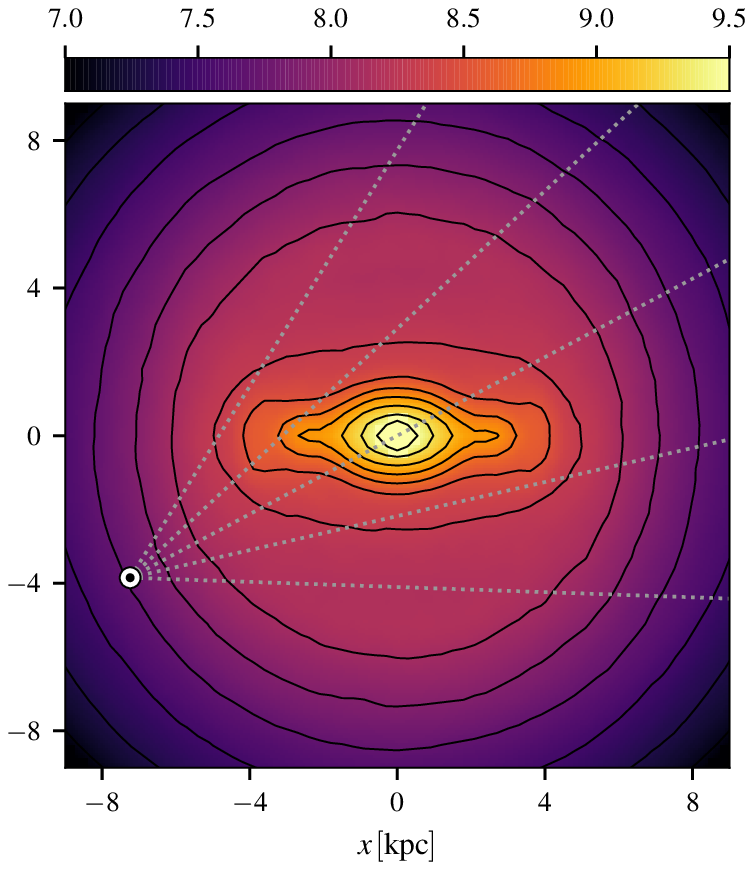}
  \includegraphics[width=0.52\linewidth]{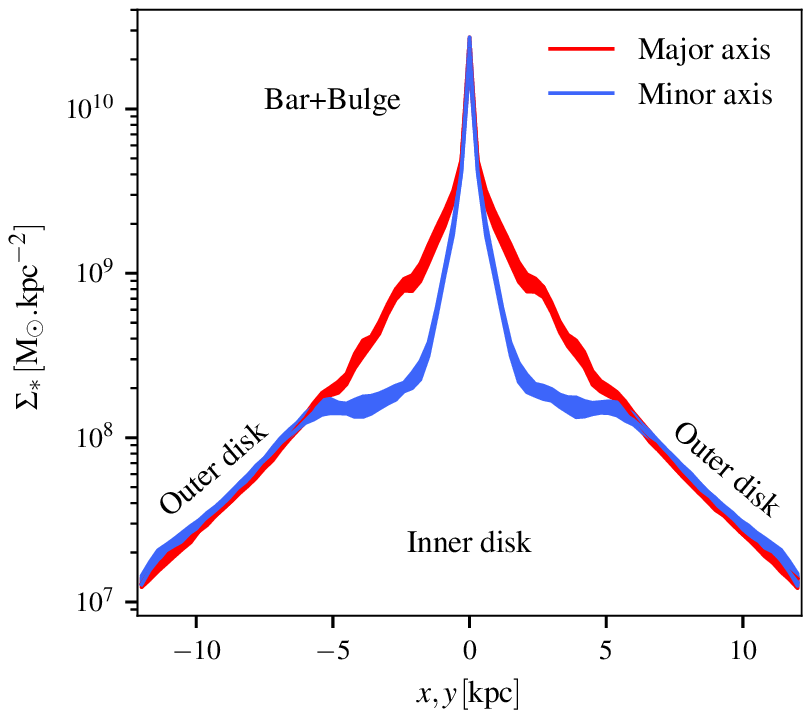}
  \vspace*{-0.25 cm}
  \caption{Face-on stellar surface density of the best Bulge and Bar model of
    P17a. The position of the Sun and sightlines for longitudes
    $l=-30\dg,-15\dg,0,15\dg,30\dg$ are indicated. }
  \label{f1SD}
  \vspace*{-0.15 cm}
  \caption{Model surface density profiles along the major and in-plane minor axis of the
    Galactic Bar.  Both figures adapted from \citet{Portail2017a}. }
  \label{f2SDP}
  \vspace*{-0.15 cm}
\end{center}
\end{figure}

\par\noindent {\textsfbi{Pattern speed}}.  P17a constructed M2M models for different values
of the bar pattern speed $\Omega_b$, the mass per RCG star, and the mass of the central
nuclear stellar disk (NSD). The NSD influences in particular the central $\sigma_r$-profile
and the vertical proper motions, which are constrained from the BRAVA and OGLE surveys, and
from these data, its mass must be $\sim2\times10^9\Msun$.

Good fits to the kinematic observables are obtained for a range of $\Omega_b$. Joint
$\chi^2$ for the BRAVA and ARGOS data and a systematic error estimate gives
$\Omega_b=39\pm3.5\kmskpc$. The pattern speed influences both the mean rotation and
dispersion in the Bulge, while the mass in the bulge region influences only
$\sigma_r$. Hence $\Omega_b$ and the mass can both be determined by the bulge kinematics. In
the future, kinematics in the long bar region is expected to yield an independent
constraint on $\Omega_b$. The value found from the bulge stellar kinematics is in good
agreement with recent analysis of MW gas dynamics \citep{Sormani2015}.  With a bar
half-length $R_b\narreq5\pm0.2\kpc$, the derived pattern speed corresponds to a corotation
radius $R_{\rm cr}=6.1\pm0.5\kpc$ and ${\cal R}\!\equiv R_{\rm cr}/R_b\narreq1.2\pm0.1$,
which is conventionally described as a fast bar.

\newtopic {\textsfbi{The Hercules stream - stars on Lagrange orbits visit the Sun}}.  The
Hercules stream is a substantial kinematic subgroup in the solar neighbourhood (SNd)
$(U,V)$-distribution with negative $U\narrsim-30\kms$ (moving outward) and negative
$V\narrsim-50\kms$ (slower than mean rotation) relative to the Sun. In its conventional
interpretation, is is identified with outer Lindblad resonance (OLR) orbits of the bar
\citep{Dehnen2000, Antoja2014}, whose corotation radius is then $R_0/R_{\rm cr}=1.83\pm0.02$
($\Omega_b=53\pm0.5\kms$ for $R_0,V_0$ given above). This would place corotation clearly
within the bar. The OLR explanation is incompatible with the best models obtained by P17a
from bulge-bar data which have OLR at $\sim10.5\kpc$ radius.

\begin{figure}[t]
\vspace*{-0.1 cm}
\begin{center}
  \includegraphics[width=0.47\linewidth]{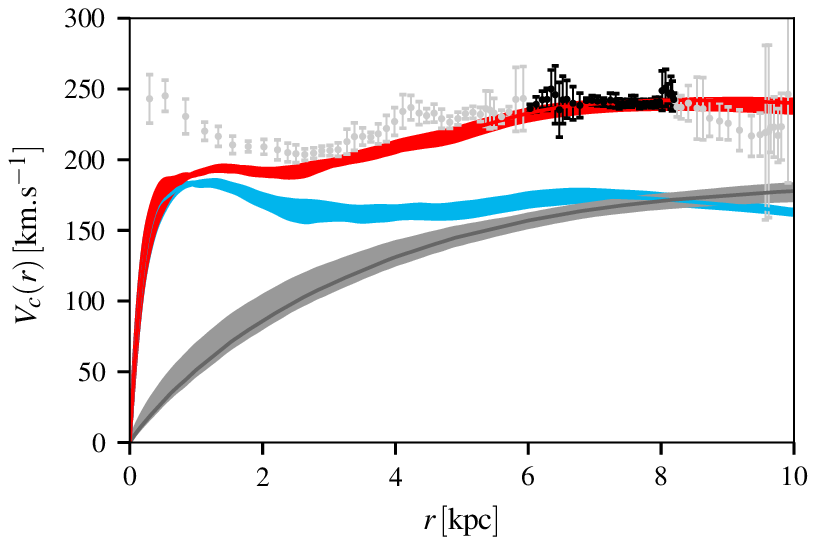}
  \includegraphics[width=0.49\linewidth]{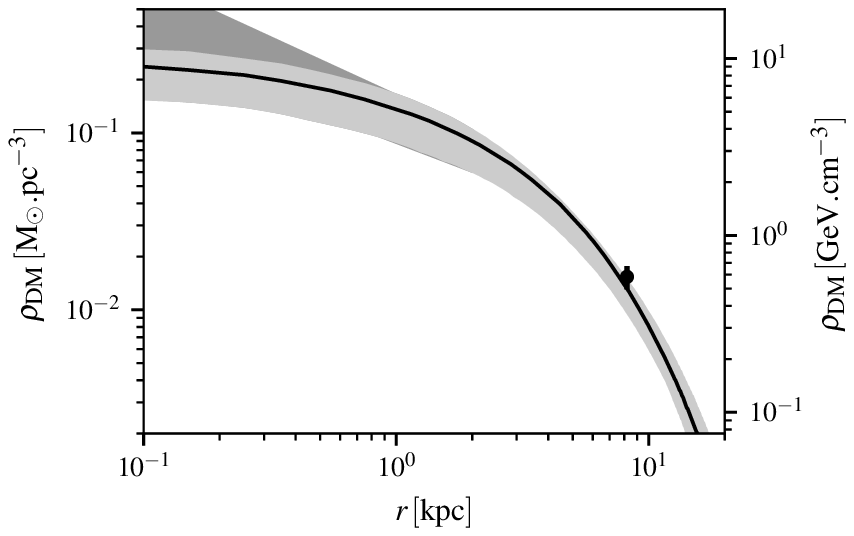}
%% \vspace*{-1.0 cm}
  \caption{MW rotation curve based on terminal velocities and $V_0=238\kms$. Less reliable
    velocities, because likely influenced by bar-induced non-circular streaming velocities,
    are shown in grey. Coloured bands show contributions from stars (blue), dark halo
    (grey), and total (red).}
  \label{f3RC}
  \vspace*{-0.1 cm}
  \caption{Range of dark matter halo profiles from systematic variation models (light
    grey) around best-fitting model of P17a.  Range of possible inner slope variations
    shown in dark grey. The models require a central $\sim 2\kpc$ core to simultaneously
    account for the low dark matter fraction in the Bulge and the rotation curve near
    $R_0$. The data point at 8.2 kpc is the local value from \citet{Piffl2014} which was
    not fitted. Figures adapted from \citet{Portail2017a}.  }
  \label{f4HD}
  \vspace*{-0.2 cm}
\end{center}
\end{figure}

Although not made for the SNd, these models however suggest an alternative explanation,
without any additional fitting. \citet{Perez-Villegas2017} found a kinematic subgroup much
like Hercules in the favoured P17a model due to stars orbiting the Lagrange points
of the bar and visiting the SNd.  These orbits extend from inside corotation to just outside
the solar radius, predicting naturally that the Hercules stream is more prominent inwards
from the Sun and nearly absent further out, and that these stars might be older and more
metal-rich than other stars near the Sun. This hypothesis can soon be tested with Gaia-DR2
data which will show us the combined effect of the bar and spiral arms on the kinematics in
the SNd (the latter were not included by P17a).

\newtopic {\textsfbi{Stellar mass distribution}}.  Figure~\ref{f1SD} shows the face-on
stellar mass distribution of the inner MW obtained from the P17a dynamical model. The
bulge and long bar density are essentially determined by the starcounts. The inner disk
around the Bar connects in a dynamically self-consistent way the stellar surface density
rising inwards from the Sun with the steeply falling density on the minor axis of the
Bulge; direct constraints by the data on the inner disk density are still
weak. Figure~\ref{f2SDP} shows the surface density profiles along the bar major and minor
axes obtained by the model. While along the bar major axis the inward exponential rise
continues along the Bar into the Bulge, the disk surface density along the minor axis is
essentially flat between the corotation radius and the Bulge. Such density structures are
known from external galaxies, e.g., the bar-lens galaxy NGC 4314 \citep{Laurikainen2014}.

Stellar masses measured by the model for the photometric Bulge and Bar, the Bulge alone, and
for the inner Disk ($R<5.3\kpc$) are $M_{bb}=1.9\times10^{10}\Msun$, $M_{\rm
  bulge}=1.3\times10^{10}\Msun$, and $M_{id}=1.3\times10^{10}\Msun$, respectively, with
typical uncertainties $\sim0.1\times10^{10}\Msun$. These values depend on the mass-to-RCG
ratio, taken as $M/N_c=1000\pm100\Msun$/RCG star in P17a. This number was obtained from
relating the stellar mass in an HST fields to the surface density of RCG counts nearby,
analogous to the method of \citet{Valenti2016} but taking into account the variation of
the red giant background with latitude. \citet{Wegg2017} show that the microlensing
time-scale distribution fitted with a stellar mass function derived from a three-power law
IMF is very similar to a Kroupa IMF, and results in $M/N_c=960\pm100\Msun$/RCG. Indeed, the
P17a model gives an excellent a posteriori fit also to the microlensing optical depth which
is a measure of the integrated stellar surface density between the Sun and the source stars
in the Bulge.

\begin{figure}[t]
%\vspace*{-0.3 cm}
\begin{center}
  \includegraphics[width=0.98\linewidth,viewport=0 220 794 585,clip=true]{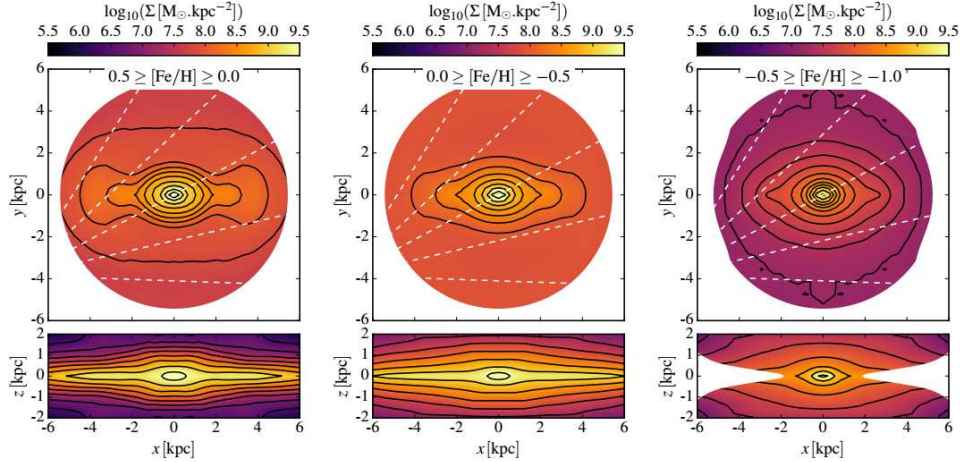}
%% \vspace*{-1.0 cm}
  \caption{Face-on and edge-on surface density distributions of inner MW stars in three
    metallicity bins.  The strong dependence of the orbit distribution on metallicity is
    apparent. Adapted from \citet{Portail2017b}.}
  \label{f5MC}
\end{center}
\end{figure}

\newtopic {\textsfbi{The core in the MW's dark matter profile}}.  The total dynamical mass
in the VVV box (see above) is very well determined in these models,
$M_{dy}=1.85\pm0.05\times10^{10}\Msun$ including systematic uncertainties from variations
in data and model assumptions. \citet{Portail2015} had previously found
$M_{dy}=1.85\pm0.05\times10^{10}\Msun$, entirely compatible.  Because also the stellar
mass in the bulge region is well-determined, the dark matter mass in the VVV box cannot
exceed $\sim0.3\times10^{10}\Msun$. Figure~\ref{f3RC} shows the MW rotation curve (RC),
based on terminal velocities and $V_0$. This fixes the dark matter mass at $R_0$ and its
profile in the $6\nto8\kpc$ range. In order to match the low dark matter fraction in the
bulge, the dark matter profiles must then flatten to a core or shallow cusp at
$\sim2\kpc$; see Figure~\ref{f4HD}.

\section{Chemo-Dynamics: The multi-component bulge}

\par\noindent {\textsfbi{Metallicity-dependent kinematics and MDF in the bulge}}.  The
near-cylindrical rotation of bulge stars is seen for all metallicities up to
$\rm{[Fe/H]}\narrsim-1$ in the ARGOS survey, and was comfirmed by APOGEE and GIBS also at
low $|b|$. The ARGOS, GIBS, and GES \citep{Rojas-Arriagada2017} surveys show distinct
velocity dispersion properties between the metal-rich and metal-poor bulge stars.  The
metal-rich component has a steep gradient with $|b|$ while the metal-poor component has
flatter dispersion profiles with both $|l|$ and $|b|$. Both components have roughly equal
$\sigma_r$ at $|b|\narreq2\dg$, and the metal-rich component has higher (lower) dispersion
at lower (higher) latitudes.

While the GIBS and GES surveys find two components in their MDF, with different
kinematics, ARGOS finds three main bulge and additional metal-poor components in the MDF
which all differ in their kinematics. These differences are likely to be due at least in
parts to the different SSF: while the ARGOS SSF gives high weight to the outer bulge
\citep{Freeman2013}, the GIBS SSF includes mostly stars within $\pm0.6\kpc$ around the
peak density \citep{Zoccali2014}, and the GES SSF around $\sim\!\pm2.3\kpc$
\citep{Rojas-Arriagada2017}. However, survey cross-validation by observing/analysing the
same stars would be important in order to check for non-SSF related differences in the
MDF.

\begin{figure}[t]
\vspace*{-0.3 cm}
\begin{center}
  \includegraphics[width=0.48\linewidth]{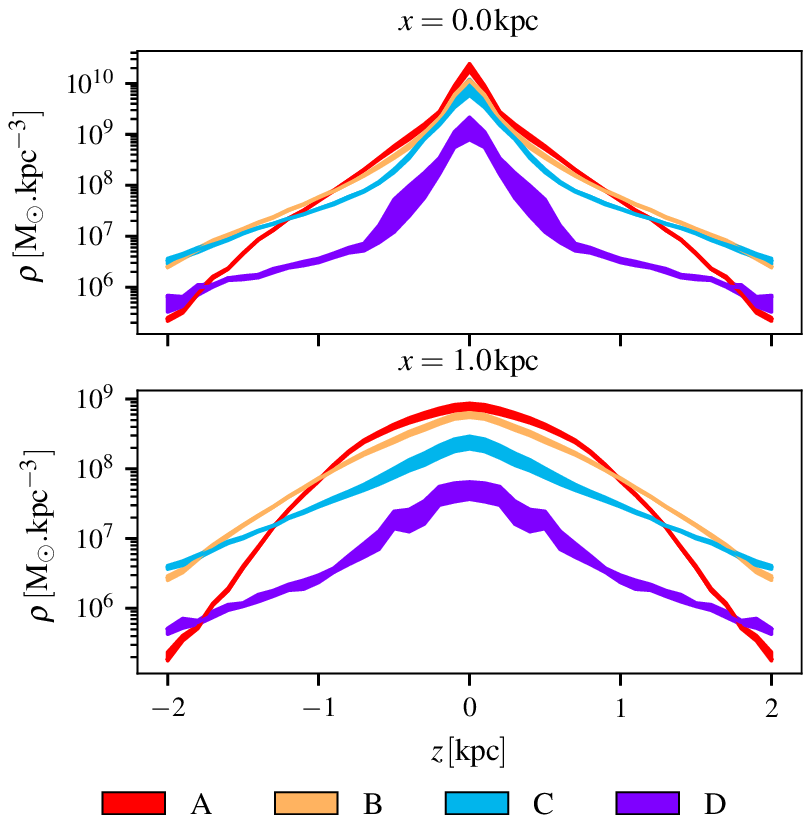}
  \includegraphics[width=0.48\linewidth,viewport=0 -20 240 240]{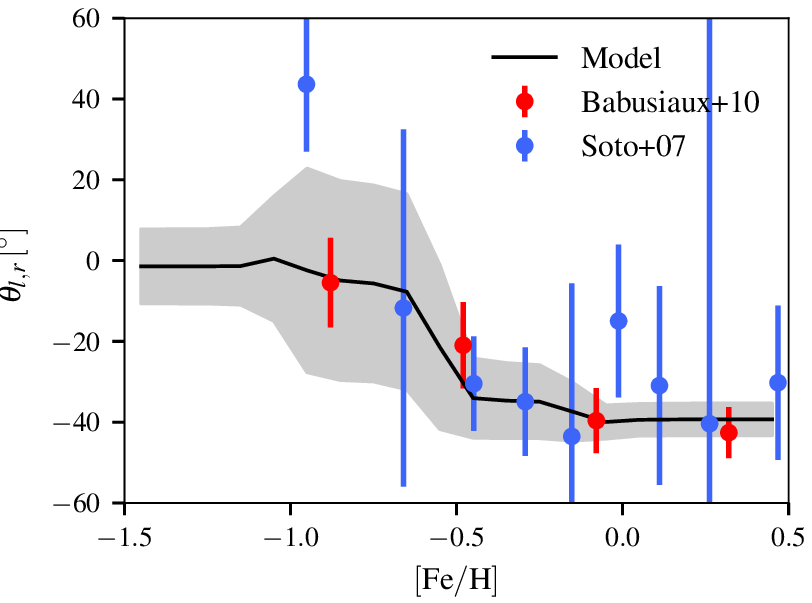}
  \caption{Vertical density profiles for Bulge stars in different metallicity bins, at the
    centre and at distance $x=1\kpc$ down the Bulge major axis. Adapted from
    \citet{Portail2017b}. }
  \label{f6VPR}
  \caption{Vertex deviation of stars in Baade's Window \citep{Soto2007, Babusiaux2010},
    with the P17b model and uncertainty range overplotted. Uncertainties are large where
    the velocity ellipsoid is nearly spherical. Adapted from \citet{Portail2017b}.}
  \label{f7VD}
\end{center}
\end{figure}

\newtopic {\textsfbi{Chemo-dynamical equilibrium (CDE) models with M2M}}.
\citet[][hereafter P17b]{Portail2017b} extended the M2M method to an augmented phase-space
including metallicity, such that all particles carry information about $x$, $v$, and the
MDF.  The particle MDFs are parametrized by a multi-Gaussian expansion with individual
Gaussians adjusted to the ARGOS metallicity bins. Particles are projected into observational
space using isochrones and metallicity-dependent selection functions taking account of the
survey SSF.  Their metallicity weights are adjusted by comparing to the number, mean
velocity, and LOS velocity dispersion of stars from the ARGOS and APOGEE surveys in bins of
distance and metallicity, while following their orbits in the overall barred gravitational
potential from P17a.  When fitted to the data, the final CDE model returns an interpolation
for the distribution of stars over orbits and metallicity, or over $x$, $v$, and
metallicity. From this one can reconstruct the 3D density, kinematics, and orbital structure
of stellar populations in different metallicity bins.

\newtopic {\textsfbi{The metal-rich bar components}}.  Figure~\ref{f5MC} based on P17b
shows face-on and edge-on surface densities for stars with $\mathrm{[Fe/H]} \ngt 0$
(metal-rich bin A), $-0.5\nlt\mathrm{[Fe/H]}\nle 0$ (intermediate bin B), and
$-1.0\nlt\mathrm{[Fe/H]}\nle-0.5$ (metal-poor bin C). In order of decreasing metallicity,
the photometric selection of the ARGOS sample leads to $(23\%, 43\%, 29\%)$ of ARGOS stars
in the bins (A, B, C), while the final SSF-corrected CDE model of the bar region has
$(52\%, 34\%, 12\%)$ in bar-supporting and $(38\%, 47\%, 14\%)$ in not-bar supporting
orbits in these bins, showing the importance of the SSF. Since estimated metallicity
errors are $\sim\!0.1$ dex \citep{Ness2013a}, much smaller than the total metallicity
range, uncertainties in the MDF can only have small effects on the metallicity
ordering. The figure illustrates that indeed stars in all metallicity bins are
significantly barred.  In terms of mass, most of the support to the bar is is provided by
the metal-rich stars. Bin A stars contribute most to the Galactic bar and B/P bulge; they
have dynamical properties consistent with a disk origin. Stars in bin B are hotter and
rotate slightly faster than stars in A, they are more extended vertically, and contribute
somewhat less to the bar and B/P shape. They are consistent with a disk origin formed from
stars located initially at larger radii \citep{DiMatteo2014}.

\newtopic {\textsfbi{The metal-poor thick disk-like stars}}.  Metal-poor stars in bin C
($\mathrm{[Fe/H]} \leq -0.5$) rotate slower and have higher dispersion than the more metal
rich stars. As shown in Fig.~\ref{f5MC}, they contribute weaker support to the bar and do
not support the B/P shape.  Figure~\ref{f6VPR} shows that outside the central $\kpc$, these
metal-poor stars are found to have the density distribution of a thick disk bar; in these
regions their vertical profile is exponential with scale height $\narrsim 500$ pc. These stars
also show cylindrical rotation \citep[][P17b]{Ness2013b}, further confirming their thick
disk nature.

However, in the inner ($x\lta1\kpc$, $z\lta0.6\kpc$) of the Bulge, Fig.~\ref{f5MC} shows
evidence for an extra component of these metal-poor stars with short scale height rising
towards the Galactic center; this component is seen also for stars in the even more
metal-poor bin D. See also \citet{Pietrukowicz2015} for RR Lyrae stars and
\citet{Zoccali2017} for stars in GIBS. This component could consist of thick disk stars on
orbits compressed by the deep gravitational potential of the nuclear disk, stars from the
inner halo-bulge, or stars from a (so far unconfirmed) classical bulge formed by early
mergers.

The combined orbit distributions of all metallicities in the model of P17b naturally
reproduce the observed vertex deviations in Baade's window, see Figure~\ref{f7VD}. The
absence of a significant vertex deviation for stars in this plot with [Fe/H]$<-0.5$ has
been interpreted as a signature of an old classical bulge by, e.g.,
\citet{Soto2007}. However, the stars with [Fe/H]$<-0.5$ in Baade's Window are
predominantly from the thick disk-bar component of bin C; the lack of significant vertex
deviation for these metal-poor stars is caused by this thick disk-bar distribution and
does not imply a large classical bulge component.

\section{Conclusions and outlook}

\par\noindent {\textsfbi{Conclusions}}. We live in a strongly barred galaxy with a
predominant B/P bulge. The bar region contains 2/3 of the MW’s stellar mass. The rotation
curve and the low dark matter fraction in the bulge require a $\sim2\kpc$ core in the Galaxy's
dark matter halo.  Different stellar populations in the bulge have clearly different orbit
distributions.  This must be exemplary for most other bulge-like stellar systems, and make
MW studies relevant for galaxy studies in general.

\newtopic{\textsfbi{Outlook}}. We can look forward to the results of on-going and
near-future ground-based surveys and of the Gaia mission. Based on these data we expect to
reach an understanding of the MW's stellar populations and formation history that is unique
to our Galaxy.

On the structure of the inner Galaxy, we need to learn more about (1) the density,
kinematics, stellar population mix of the nuclear disk, inner disk, and long bar, (2) the
properties of the spiral arms and their masses, (3) the structure of the outer bulge and
its transition to the inner halo, and (4) the dark matter distribution in the inner
Galaxy.  On the subject of stellar populations we need to understand better (5) how many
stellar populations are there in the bulge and are they discrete or not?  (6) what are the
stellar populations in the long bar and how do they relate to those in the inner disk?
(7) is the old, metal-poor component traced by RRL related to the early stellar halo? Is
the metal-poor central concentration related to a classical bulge?  Dynamical models will
help in understanding many of these issues because they can relate stars to their orbits,
which is a lower-dimensional and easier-to-understand space than positions and velocities.

\newtopic{\textsfbi{Acknowledgments}}: I am grateful to Angeles P{\'{e}}rez-Villegas,
Matthieu Portail, and Chris Wegg for our fruitful collaboration over the last several years.

% {30 lines = 0.65pp = 5.75}

\end{document}